# FINITE-TEMPERATURE MICROMAGNETICS OF HYSTERISIS FOR MISALIGNED SINGLE IRON NANOPILLARS

M. A. Novotny[1,2], S. M. Stinnett[1,2], G. Brown[3,4], P. A. Rikvold[4,5]
[1] ERC Center for Computational Sciences, Mississippi State Univ., Mississippi State, MS, USA
[2] Department of Physics and Astronomy, Mississippi State Univ., Mississippi State, MS, USA
[3] Center for Computational Sciences, Oak Ridge National Lab, Oak Ridge, TN, USA
[4] School for Computational Science and Information Technology, Florida State University, Tallahassee, FL, USA
[5] Center for Materials Research and Technology and Department of Physics, Florida State University, Tallahassee, FL, USA

Abstract: We present micromagnetic results for the hysteresis of a single magnetic nanopillar that is misaligned with respect to the applied magnetic field. We provide results for both a one-dimensional stack of magnetic rotors and of full micromagnetic simulations. The results are compared with the Stoner-Wohlfarth model.

## 1. INTRODUCTION

Although hysteresis in single-domain nanomagnets has been known for many decades, there is currently much interest in looking anew at this phenomenon. This is partly driven by recent experiments on single-domain nanomagnets, in which the hysteresis and magnetization reversal of a single single-domain nanomagnet can be measured. It is also driven by the applications of nanomagnets, in particular to magnetic recording.

Finally, one is no longer confined to pencil-and-paper calculations to understand physical phenomena. Available computer resources allow calculations which are much more complicated. Here we present large-scale computer simulations of hysteresis for two different model systems of single-domain nanoscale Fe pillars. We focus on the hysteresis when the long axis of the pillars is misaligned with the applied magnetic field.

## 2.     METHODS AND MODELS

### 2.1     Coherent Rotation

Given a single-domain particle with uniaxial anisotropy, it is possible to find the quasi-static equilibrium position of the magnetization when a magnetic field is applied at some angle to the easy axis. It is assumed that the magnetization can be represented by a single vector, **M**, with constant amplitude, $M_S$. The energy density is then

$$E = K \sin^2\theta - M_S H \cos(\phi - \theta), \quad (1)$$

where $K$ is the uniaxial anisotropy constant, $H$ is the magnetic field applied at an angle $\phi$ to the easy axis and $\theta$ is the angle the magnetization makes with the easy axis. Stoner and Wohlfarth showed that the critical transition curve for the coherent reversal of the magnetization is given by,[1]

$$h_{AX}^{2/3} + h_{AY}^{2/3} = 1 \quad (2)$$

where $h_{AX}$ and $h_{AY}$ are the components of the magnetic field along the easy and hard axes respectively. Equation (2) is the well-known equation of a hypocycloid of four cusps, also known as an astroid.

### 2.2     Micromagnetics

For systems where the spins are not aligned and/or where the field is changing too rapidly for the magnetization to reach its quasi-static equilibrium position, it is usually necessary to use micromagnetics to determine the reversal process. The basic approach is to discretize the system into a series of coarse-grained magnetization vectors, $\mathbf{M}(\mathbf{r_i})$, where $\mathbf{r_i}$ is the position of the *i*-th magnetization vector. Each spin is assumed to have uniform magnetization, $M_S$, corresponding to the saturation magnetization of the bulk material, a valid assumption for temperatures well below the Curie temperature.[2] The time evolution of each spin is given by the Landau-

Lifshitz-Gilbert (LLG) equation,[3,4]

$$\frac{d\mathbf{M}(\mathbf{r_i})}{dt} = \frac{\gamma_0}{1+\alpha^2}\left(\mathbf{M}(\mathbf{r_i}) \times \left[\mathbf{H_T}(\mathbf{r_i}) - \frac{\alpha}{M_S}(\mathbf{M}(\mathbf{r_i}) \times \mathbf{H_T}(\mathbf{r_i}))\right]\right) \quad (5)$$

where $\mathbf{H_T}(\mathbf{r_i})$ is the total local field at the *i*-th position, $\gamma_0$ is the gyromagnetic ratio ($1.76 \times 10^7$ rad/Oe-s), and $\alpha$ is a dimensionless phenomenological damping term which determines the rate of energy dissipation. The first term represents the precession of each spin around the local field, while the second term is a dissipative term that drives the motion of the magnetization towards equilibrium. For the sign of the undamped precession term, we follow the convention of Brown.[3]

The total local field, $\mathbf{H_T}(\mathbf{r_i})$, may include contributions from the applied field (Zeeman term), the crystalline anisotropy (set to zero in our model), the dipole field, and exchange interactions. At nonzero temperatures, thermal fluctuations also contribute a term to the local field in the form of a stochastic field which is assumed to fluctuate independently for each spin. The fluctuations are assumed Gaussian, with zero mean and (co)variance given by the fluctuation- dissipation theorem.[5]

While the stochastic thermal field requires careful treatment of the numerical integration in time, the most computationally intensive part of the calculation involves the dipole term. For systems with more than a few hundred spins, it is necessary to use a more advanced algorithm. We use the Fast Multipole Method (FMM), the implementation of which is discussed elsewhere.[5]

In this paper, we examine two model systems. The first is a nanopillar with dimensions of 5.2 nm x 5.2 nm x 88.4 nm. The cross-sectional dimensions are small enough (about 2 exchange lengths) that the assumption is made that the only significant inhomogeneities in the magnetization occur along the long axis[6] (*z*-direction). The particles in this model, discussed previously,[5] are therefore discretized into a linear chain of 17 spins along the long axis of the pillar.

The second model system consists of a single nanopillar with dimensions 9 nm x 9 nm x 150 nm. The dimensions were chosen to correspond to arrays of Fe nanopillars fabricated by Wirth, *et al.*[7] In this model, the system is discretized into 4949 sites (7 x 7 x 101) on the computational lattice.

Material properties in both systems were chosen to correspond to bulk Fe. The saturation magnetization is 1700 emu/cm$^3$ and the exchange length (the length over which the magnetization can change appreciably) is 3.6 nm.

We take $\alpha = 0.1$ to represent the underdamped behavior usually assumed to be present in nanoscale magnets.

## 3.     RESULTS AND ANALYSIS

In Fig. 1 we present hysteresis loops at $T = 100$ K for the second model for the field at $0°$ and $45°$ to the long axis of the particle. The loops were calculated using a sinusoidal field of period 15 ns which started at the maximum value of 5000 Oe. In all loops shown here, the reported magnetization is the component along the long axis of the particle. At $45°$, the magnetization vector is initially pulled away from the easy axis by the large magnetic field. As the field is reduced to zero, the magnetization relaxes towards the easy axis, reaching essentially saturation at zero field.

Figure 2 shows the *z*-component of the magnetization in these pillars for various times during the switching process with the field at $45°$ to the long axis of the pillar, under the same conditions as Fig. 1. Note that the magnetic end-caps are the sources of the finite-temperature nucleation that leads to the reversal of the hysteresis. Furthermore, note that these particles do not have a uniform magnetization vector, even though they are single-domain particles.

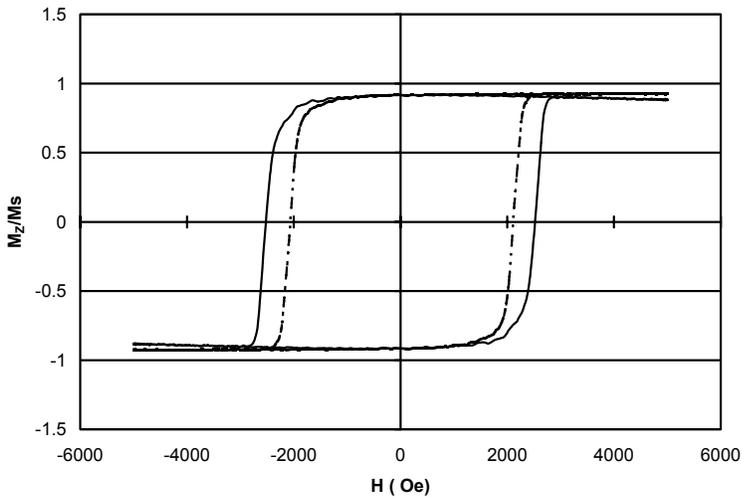

*Fig. 1 Hysterisis loops at T= 100 K for the field aligned (dashed) and misaligned at 45° (solid). The simulation was performed over one half of the loop and the results reflected to generate the entire loop.*

*Finite-Temperature Micromagnetics of Hysterisis for Misaligned Single Iron Nanopillars*  <span style="float:right">5</span>

Figure 3a presents hysteresis loops for the first model for the field misaligned at various angles. The applied field is again sinusoidal with a period of 200 ns. Note that when the applied field is perpendicular to the long axis of the pillar, the hysteresis loop has a bubble shape qualitatively consistent with both the Stoner-Wohlfarth (SW) model (Fig. 3b) and the experimental results of Wirth *et al.*[7] Quantitatively, however, the field at which nonzero magnetization appears is significantly lower than both what is expected from SW and from the experimental value.

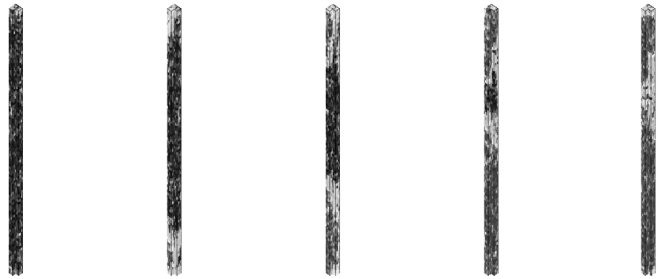

*Fig. 2 Snapshots of the z-component of the magnetization at (from left to right) 0.00ns, 4.875 ns , 5.000 ns, 5.075 ns, and 5.100 ns. One quarter of the pillar has been removed from the illustration to show the behavior of $M_Z$ inside the pillar.*

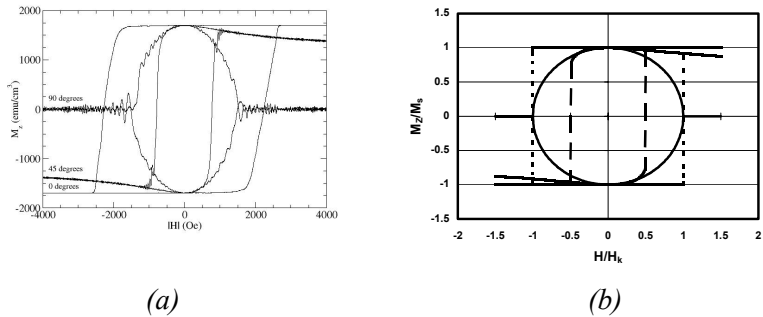

*(a)    (b)*

*Fig. 3 Hysteresis Loops for $0°$, $45°$, and $90°$ misalignment of the field with the long axis of (a) small pillars and (b) the SW model.*

## 4.   DISCUSSION AND CONCLUSIONS

In this brief paper, we presented finite-temperature micromagnetic simulations of hysteresis for a single Fe nanopillar. Two different micromagnetic models were simulated, both of which used fast periods of the sinusoidal applied field. The results were compared with the quasi-static, zero-temperature predictions of the Stoner-Wohlfarth model. We observed that when the field is aligned with the long axis of the particle, all models have qualitatively similar hysteresis loops, although the micromagnetic loops for the aligned case are more rounded than that of the SW model. When the applied field is at 45º to the long axis of the particle, the SW and small pillar models give switching fields which are lower than when the field is aligned, while the large pillar model gives a switching field which is larger than the aligned case, more consistent with experimental results.[7] When the field is perpendicular to the long axis, the hysteresis loop has a bubble shape. However, the SW model gives a switching field which is equal to the aligned case, while the small and large pillar models give switching fields that are smaller and larger than the aligned case, respectively.

These results demonstrate that although simple models of magnetization switching in magnetic nanopillars can yield qualitatively correct results, to obtain quantitatively correct results, large-scale micromagnetic calculations are required. Only such detailed calculations will enable quantitative comparisons with finite-temperature, fast-period hysteresis loops.

## ACKNOWLEDGEMENTS

The authors gratefully acknowledge support from NSF grant DMR0120310.